\documentclass{revtex4}
\usepackage{graphicx}
\usepackage{fancyhdr}
\usepackage{amsmath}
\usepackage{slashed}
\usepackage{color}

\begin{document}

\title{Examining custodial symmetry in the Higgs sector of Georgi-Machacek model}

%

\author{Cheng-Wei Chiang}
\affiliation{Department of Physics and Center for Mathematics and Theoretical Physics,
National Central University, Chungli, Taiwan 32001, ROC}
\affiliation{Institute of Physics, Academia Sinica, Taipei, Taiwan 11529, ROC}
\affiliation{Physics Division, National Center for Theoretical Sciences, Hsinchu, Taiwan 30013, ROC}

\begin{abstract}
We study consequences of custodial symmetry in the Higgs sector of the Georgi-Machacek (GM) model.  We discuss how the 5-plet and 3-plet Higgs bosons classified under the custodial SU(2) symmetry are produced and decay at the CERN Large Hadron Collider.  By determining their masses through proposed channels, we can test the mass degeneracy in each Higgs multiplet.
\end{abstract}

\maketitle

\thispagestyle{fancy}


\section{Introduction}

Since its introduction, the standard model (SM) has been successfully explaining many elementary particle phenomena for over four decades.  All particles in the model except for the Higgs boson had been discovered and studied at colliders in the past.  In the last summer, a boson of mass about 125 GeV and properties similar to the SM Higgs boson has been found at the CERN LHC.  An immediate task is to test whether this boson is indeed the one responsible for the electroweak symmetry breaking (EWSB) in the SM.  An ensuing question is whether this SM-like Higgs boson actually belongs to a larger Higgs sector than that of the SM.

In recent years, interest in models with Higgs triplet fields has been revived partly because it provides a mechanism to generate neutrino mass and partly due to its rich Higgs phenomenology.  Such models possess such novel features as lepton number violating (or even lepton flavor violating) processes and doubly charged Higgs bosons.  In the simplest version where only one complex Higgs triplet field is added to the SM~\cite{typeII}, the model suffers from the constraint of electroweak $\rho$ parameter so that the triplet vacuum expectation value (VEV) cannot be larger than a few GeV.  In this case, the charged Higgs bosons dominantly decay into a pair of leptons.  The corresponding collider phenomenology has been extensively studied in the past~\cite{HTMll}.

When an additional real Higgs triplet field is added to the above-mentioned model, as introduced by Georgi and Machacek (GM)~\cite{GM}, and a VEV alignment is assumed, the $\rho$ parameter can be maintained at unity at tree level.  It had been explicitly shown that such a VEV alignment could be achieved with a suitable Higgs potential~\cite{Chanowitz}.  One-loop radiative corrections and renormalization were studied, and the $\rho$ parameter was found to be shielded at the same level as the SM~\cite{GVW2}.  In the GM model, the triplet VEV can be comparable to the SM doublet VEV.  The scenario of large triplet VEV ($\agt$ a few tens of GeV) leads to different phenomenology because the charged Higgs bosons decay dominantly into weak gauge bosons \cite{GVW1}.  In this article, we review a recent study that analyzes the possibility of experimentally verifying consequences of custodial symmetry of the GM model~\cite{Chiang:2012cn}.

\section{Georgi-Machacek Model}

The Higgs sector of the GM model is comprised of the SM isospin doublet Higgs field $\phi$ with hypercharge $Y=1/2$ and two isospin triplet Higgs fields $\chi$ with $Y=1$ and $\xi$ with $Y=0$.  These fields can be expressed in the form:
\begin{align}
\Phi=\left(
\begin{array}{cc}
\phi^{0*} & \phi^+ \\
\phi^- & \phi^0
\end{array}\right),\quad 
\Delta=\left(
\begin{array}{ccc}
\chi^{0*} & \xi^+ & \chi^{++} \\
\chi^- & \xi^0 & \chi^{+} \\
\chi^{--} & \xi^- & \chi^{0} 
\end{array}\right), \label{eq:Higgs_matrices}
\end{align}
where $\Phi$ and $\Delta$ are transformed under $SU(2)_L\times SU(2)_R$ as $\Phi\to U_L\Phi U_R^\dagger$ and $\Delta\to U_L\Delta U_R^\dagger$ with $U_{L,R}=\exp(i\theta_{L,R}^aT^a)$ and $T^a$ being the corresponding $SU(2)$ generators.  The neutral components in Eq.~(\ref{eq:Higgs_matrices}) can be parametrized as 
\begin{align}
\phi^0=\frac{1}{\sqrt{2}}(\phi_r+v_\phi+i\phi_i), \quad 
\chi^0=\frac{1}{\sqrt{2}}(\chi_r+i\chi_i)+v_\chi,\quad
\xi^0=\xi_r+v_\xi, \label{eq:neutral}
\end{align}
where $v_\phi$, $v_\chi$ and $v_\xi$ are the VEV's for $\phi^0$, $\chi^0$ and $\xi^0$, respectively.  Assuming $v_\chi=v_\xi \equiv v_\Delta$, the $SU(2)_L\times SU(2)_R$ symmetry is reduced to custodial $SU(2)_V$ symmetry.  The phase convention for the component scalar fields are chosen to be the same as in Ref.~\cite{Chanowitz}.  The relevant Lagrangian involving the Higgs fields is
$\mathcal{L}_{\text{GM}}
=\mathcal{L}_{\text{kin}}+\mathcal{L}_{Y}+\mathcal{L}_{\nu}-V_H$, 
where $\mathcal{L}_{\text{kin}}$, $\mathcal{L}_{Y}$, $\mathcal{L}_{\nu}$ and $V_H$ 
represent kinetic term, Yukawa interactions between $\phi$ and the fermions, Yukawa interactions between $\chi$ and the lepton doublets, and the Higgs potential, respectively. 

The most general Higgs potential invariant under the $SU(2)_L\times SU(2)_R\times U(1)_Y$ symmetry is
\begin{align}
V_H&=m_1^2\text{tr}(\Phi^\dagger\Phi)+m_2^2\text{tr}(\Delta^\dagger\Delta)
+\lambda_1[\text{tr}(\Phi^\dagger\Phi)]^2
+\lambda_2[\text{tr}(\Delta^\dagger\Delta)]^2
+\lambda_3\text{tr}[(\Delta^\dagger\Delta)^2]
+\lambda_4\text{tr}(\Phi^\dagger\Phi)\text{tr}(\Delta^\dagger\Delta) \notag\\
&+\lambda_5\text{tr}\left(\Phi^\dagger\frac{\tau^a}{2}\Phi\frac{\tau^b}{2}\right)
\text{tr}(\Delta^\dagger t^a\Delta t^b)
+\mu_1\text{tr}\left(\Phi^\dagger \frac{\tau^a}{2}\Phi\frac{\tau^b}{2}\right)(P^\dagger \Delta P)^{ab}
+\mu_2\text{tr}\left(\Delta^\dagger t^a\Delta t^b\right)(P^\dagger \Delta P)^{ab}, \label{eq:pot}
\end{align}
where $\tau^a$ are the Pauli matrices, and 
\begin{align}
t^1=\frac{1}{\sqrt{2}}\left(
\begin{array}{ccc}
0 & 1 & 0 \\
1 & 0 & 1 \\
0 & 1 & 0
\end{array}\right) ~,~
t^2=\frac{1}{\sqrt{2}}\left(
\begin{array}{ccc}
0 & -i & 0 \\
i & 0 & -i \\
0 & i & 0
\end{array}\right) ~,~
t^3=\left(
\begin{array}{ccc}
1 & 0 & 0 \\
0 & 0 & 0 \\
0 & 0 & -1
\end{array}\right) ~,~
\mbox{and}~ 
P=\left(
\begin{array}{ccc}
-1/\sqrt{2} & i/\sqrt{2} & 0 \\
0 & 0 & 1 \\
1/\sqrt{2} & i/\sqrt{2} & 0
\end{array}\right). 
\end{align}
The SM EWSB induces the triplet fields to develop a VEV $v_\Delta$ through the $\mu_1$ term in the Higgs potential.

One can use the tadpole conditions
to eliminate the parameters $m_1^2$ and $m_2^2$ in terms of the other parameters in the Higgs potential.
Moreover, we have $v^2=v_\phi^2+8v_\Delta^2=1/(\sqrt{2}G_F)$, and define the angle $\theta_H$ via $\tan\theta_H=2\sqrt{2}v_\Delta/v_\phi$.  We will use $s_H=\sin\theta_H$ and $c_H=\cos\theta_H$. 
We also introduce the parameters
\begin{align}
M_1^2=-\frac{v}{\sqrt{2}s_H}\mu_1,\quad 
M_2^2=-3\sqrt{2}s_Hv\mu_2. 
\end{align}

Under the custodial $SU(2)_V$ symmetry, the triplet field $\Delta$ can be decomposed into a 5-plet, a 3-plet, and a singlet, which are related to the original component fields as 
\begin{align}
&H_5^{\pm\pm} = \chi^{\pm\pm},\quad H_5^\pm = \frac{1}{\sqrt{2}}(\chi^\pm-\xi^\pm),\quad H_5^0=\frac{1}{\sqrt{3}}(\chi_r-\sqrt{2}\xi_r), \notag\\
&\tilde{H}_3^\pm = \frac{1}{\sqrt{2}}(\chi^\pm+\xi^\pm),\quad \tilde{H}_3^0 = \chi_i,
\quad
\tilde{H}_1^{0} = \frac{1}{\sqrt{3}} (\xi_r+\sqrt{2}\chi_r). \label{eq:cust}
\end{align}
$H_5^0$ and $\tilde{H}_1$ are CP-even states, whereas $\tilde{H}_3^0$ is a CP-odd state. 
In Eq.~(\ref{eq:cust}), the scalar fields with a tilde are not mass eigenstates in general, and can in principle mix with the corresponding scalar fields from the Higgs doublet field.  After rotating to the mass eigenstates, denoted by the symbols without a tilde, one obtain the mass eigenvalues
\begin{align}
m_{H_5}^2
&\equiv m_{H_5^{++}}^2=m_{H_5^{+}}^2=m_{H_5^{0}}^2
= \left(s_H^2\lambda_3 -\frac{3}{2}c_H^2\lambda_5\right)v^2+c_H^2M_1^2+M_2^2 ~, \notag \\
m_{H_3}^2
&\equiv m_{H_3^{+}}^2=m_{H_3^{0}}^2 
= -\frac{1}{2}\lambda_5v^2+M_1^2 ~, \notag \\
m_{H_1}^2
&\equiv m_{H_1^{0}}^2
= (M^2)_{11}s_\alpha^2+(M^2)_{22}c_\alpha^2-2(M^2)_{12}s_\alpha c_\alpha ~, \notag \\
m_{h}^2 &=(M^2)_{11}c_\alpha^2+(M^2)_{22}s_\alpha^2
+2(M^2)_{12}s_\alpha c_\alpha ~,
\end{align}
where
\begin{align}
(M^2)_{11}&=8c_H^2\lambda_1v^2 ~,~ 
(M^2)_{22}=s_H^2(3\lambda_2+\lambda_3)v^2+c_H^2M_1^2 -\frac{1}{2}M_2^2 ~, \notag \\
(M^2)_{12}&=\sqrt{\frac{3}{2}}s_Hc_H[(2\lambda_4+\lambda_5)v^2-M_1^2] ~,
\end{align}
and $c_\alpha=\cos\alpha$, $s_\alpha=\sin\alpha$, and the mixing angle $\alpha$ is defined by
$\tan2\alpha = 2(M^2)_{12} / [(M^2)_{11}-(M^2)_{22}]$ ~. 

The five dimensionless couplings in the potential, $\lambda_1, \dots, \lambda_5$, can be replaced by the five physical parameters $m_{H_5}$, $m_{H_3}$, $m_{H_1}$, $m_{h}$ and $\alpha$. 
%
The decoupling limit of this model can be obtained when we take the $v_\Delta\to 0$ limit (or equivalently $s_H\to 0$), in which the mass formulas of the Higgs bosons reduce to
\begin{align}
m_{H_5}^2 =  -\frac{3}{2}\lambda_5v^2 + M_1^2 + M_2^2,~
m_{H_3}^2 =  -\frac{1}{2}\lambda_5v^2 + M_1^2 ,~
m_{H_1}^2 =  M_1^2 -\frac{1}{2}M_2^2,~
m_h^2     = 8\lambda_1 v^2. \label{dec}
\end{align}
Consequently, the triplet-like Higgs bosons decouple when $M_1^2\gg v^2$, and 
only $h$ remains at the electroweak scale and acts like the SM Higgs boson. 
In addition, in the decoupling region $v_\Delta\simeq 0$, we find a simple mass relation for the 
triplet-like Higgs bosons:
$2 m_{H_1}^2= 3 m_{H_3}^2 - m_{H_5}^2$.

For the convenience in discussing interactions between leptons and the Higgs triplet field, 
we reorganize the Higgs fields as follows: 
\begin{align}
\phi=\left(
\begin{array}{c}
\phi^+\\
\phi^0
\end{array}\right),\quad
\chi=\left(
\begin{array}{cc}
\frac{\chi^+}{\sqrt{2}} & -\chi^{++} \\
\chi^0 & -\frac{\chi^+}{\sqrt{2}} 
\end{array}\right),\quad 
\xi=\left(
\begin{array}{cc}
\frac{\xi^0}{\sqrt{2}} & -\xi^+ \\
\xi^- & -\frac{\xi^0}{\sqrt{2}}
\end{array}\right). \label{22mat}
\end{align}
The Yukawa interactions between the lepton doublets and the Higgs triplet are
\begin{align}
\mathcal{L}_{\nu}&=h_{ij}\overline{L_L^{ic}}i\tau_2\chi L_L^j+\text{h.c.} \label{Eq:nuYukawa}
\end{align}
If we assign two units of lepton number to $\chi$, then the $\lambda_5$ and $\mu_1$ terms in the Higgs potential violate the lepton number.  If we then take $\lambda_5=\mu_1=0$, $H_3^0$ becomes massless and corresponds to the NG boson for the spontaneous breakdown of the global $U(1)$ lepton number symmetry.  In fact, $H_3^\pm$ are also massless in that case because of the custodial symmetry.
%
For simplicity, we assume that the neutrino mass eigenstates and flavor eigenstates are identical.  In terms of the scalar mass eigenstates, the interaction terms are
\begin{align}
\mathcal{L}_\nu 
&=\frac{2\sqrt{2}m_\nu}{s_Hv}H_5^{++}\overline{e_i^c}P_Le_i
-\frac{2\sqrt{2}m_\nu}{s_Hv}\left(H_5^++c_HH_3^++s_HG^+\right)\overline{\nu_i^c}P_Le_i\Big]\notag\\
&+\frac{2m_\nu}{s_Hv}\left[\frac{1}{\sqrt{3}}(H_5^0+\sqrt{2}s_\alpha h+c_\alpha H_1^0)+i(G^0 s_H +H_3^0c_H)\right]\overline{\nu_i^c}P_L\nu_i
+\text{h.c.}\label{nY}
\end{align}
On the other hand, the Yukawa interaction between the fermions of one generation and the physical Higgs bosons is
\begin{align}
\mathcal{L}_{Y}&=
-\sum_{f=u,d,e}\frac{m_f}{v}
\left[\frac{c_\alpha}{c_H}\bar{f}fh-\frac{s_\alpha}{c_H}\bar{f}fH_1^0+i\text{Sign}(f) \tan\theta_H\bar{f}\gamma_5fH_3^0\right]
\notag\\
&-\frac{\sqrt{2}V_{ud}}{v}\left[\tan\theta_H\bar{u}(m_u P_L-m_d P_R)dH_3^+\right]+\frac{\sqrt{2}m_e}{v}\tan\theta_H\bar{\nu}P_ReH_3^++\text{h.c.},  \label{Yukawa1}
\end{align}
where $V_{ud}$ is one element of the Cabibbo-Kobayashi-Maskawa (CKM) matrix, Sign$(f=u)=+1$ and Sign$(f=d,e)=-1$.

Finally, from the kinetic terms for the Higgs fields,
one obtains the same gauge boson masses as in the SM under the condition $v_\chi=v_\xi \equiv v_\Delta$. 
Thus, the electroweak rho parameter $\rho = m_W^2/(m_Z^2\cos^2\theta_W)$ is unity at the tree level.  One-loop corrections to $\rho$ have been calculated in Ref.~\cite{GVW2} for the GM model.  The Gauge-Gauge-Scalar (Gauge-Scalar-Scalar) vertices are listed in Table~III (Table~IV) in Appendix~B of Ref~\cite{Chiang:2012cn}.  One special feature is that the GM model has a tree-level $H_5^\pm W^\mp Z$ vertex.  In most Higgs-extended models with $\rho=1$ at the tree level and having singly-charged Higgs bosons ({\it e.g.}, the 2HDM), the $H^\pm W^\mp Z$ vertex is absent at the tree level~\cite{Grifols} and is only induced at loop levels and therefore much smaller than that in the GM model.  Thus, this vertex can be used to discriminate models with singly-charged Higgs bosons.  The possibility of measuring the $H^\pm W^\mp Z$ vertex has been discussed in Refs.~\cite{HWZ-LHC} for the LHC
and in Ref.~\cite{HWZ-ILC} for future linear colliders.

\section{Collider Phenomenology}

Decay branching ratios of the Higgs bosons depend on the mass parameters $m_{H_5}$, $m_{H_3}$ and $m_{H_1}$, the VEV of the triplet field $v_\Delta$, and the mixing angle $\alpha$.  The mass of the SM-like Higgs boson $h$ is fixed at 125 GeV.  We take $\Delta m \equiv m_{H_3}-m_{H_5}$, $m_{H_3}$ and $v_\Delta$ as the input parameters, and assume $\alpha=0$ for simplicity.  
Once we apply the mass relation, there are three different patterns of masses for 
the triplet-like Higgs bosons.  In the case of $\Delta m=0$, all the masses of the triplet-like Higgs bosons are degenerate: $m_{H_5}=m_{H_3}=m_{H_1}$, whereas in the case of $\Delta m>0$ ($\Delta m<0$), the mass spectrum is then $m_{H_1}>m_{H_3}>m_{H_5}$ ($m_{H_5}>m_{H_3}>m_{H_1}$).

\begin{figure}[t]
\begin{center}
\includegraphics[width=70mm]{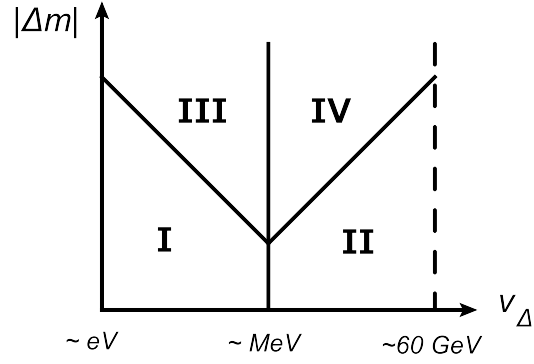}
\end{center}
\caption{Four regions with different decay patterns are schematically shown on the $v_\Delta$-$|\Delta m|$ plane.  }
\label{schematic}
\end{figure}

The decay properties of the 5-plet Higgs bosons and the 3-plet Higgs bosons can be separately considered for four different regions in the $v_\Delta$-$\Delta m$ plane, as schematically shown in Fig.~\ref{schematic}. 
In Region~I, all the triplet-like Higgs bosons mainly decay leptonically: 
\begin{align}
H_5^{++}\to \ell^+\ell^+,\quad H_5^{+}\to \ell^+\nu,\quad H_5^0\to \nu\nu,
\quad H_3^+\to \ell^+\nu,\quad H_3^0\to \nu\nu.
\end{align}
In this region, the mass of the 5-plet Higgs bosons is constrained to be $m_{H_5}\gtrsim 400$ GeV by the search at the LHC for doubly-charged Higgs bosons decaying into same-sign dileptons~\cite{LHCll}. 
In Region~II, the 5-plet Higgs bosons mainly decay into the weak gauge boson pairs, 
while the 3-plet Higgs bosons decay into the fermion pairs. 
When the mass of the 3-plet Higgs bosons is less than the top quark mass, the main decay modes are 
\begin{align}
H_5^{++}\to W^+W^+,\quad H_5^{+}\to W^+Z,\quad H_5^0\to W^+W^-/ZZ, 
\quad H_3^+\to \tau^+\nu/c\bar{s},\quad H_3^0\to b\bar{b}. \label{S2}
\end{align}
For Region~III and Region~IV, one has to separately consider the cases whether the sign of $\Delta m$ is positive or negative.  In the case of $\Delta m > 0$, the 5-plet Higgs bosons  
mainly decay into the lepton pairs (weak gauge boson pairs) in Region III (Region IV). 
The 3-plet Higgs bosons mainly decay into a 5-plet Higgs boson and a weak gauge boson: 
\begin{align}
&H_5^{++}\to \ell^+\ell^+~(W^+W^+),\quad H_5^{+}\to \ell^+\nu~(W^+Z),\quad H_5^0\to \nu\nu~(W^+W^-/ZZ),\notag\\
&H_3^+\to  H_5^{++} W^{-}/H_5^{+} Z/H_5^0 W^+,\quad 
H_3^0\to H_5^\pm W^{\mp }/H_5^0Z.
\end{align}
In the case of $\Delta m<0$, 
the main decay modes in both Region~III and Region~IV are
\begin{align}
H_5^{++}\to H_3^+ W^+,\quad H_5^{+}\to H_3^+ Z/H_3^0 W^+,\quad H_5^0\to H_3^\pm W^\mp/H_3^0 Z
\quad H_3^+\to H_1^0W^+ ,\quad H_3^0\to H_1^0Z.
\end{align}

There are several production modes for the 5-plet Higgs bosons $H_5$ and the 3-plet Higgs bosons $H_3$, as listed below.  Here $q,q',Q,Q'$ and those with bars denote light quarks and anti-quarks.

\begin{description}
\item[1.] {\it The Drell-Yan process: }
$H_5$ and $H_3$ can be produced in pairs via $\gamma$ and $Z$, {\it e.g.}, $pp\to H_5H_5$ and $pp\to H_3H_3$.  The cross section is determined by the gauge coupling as well as the Higgs masses $m_{H_5}$ and $m_{H_3}$, independent of the value of $v_\Delta$.  
\item[2.] {\it The mixed Drell-Yan (mDY) process: }
$H_5$ and $H_3$ can be produced at the same time, {\it e.g.}, $pp\to H_5H_3$, which we call the mixed Drell-Yan (mDY) process to be separated from the usual Drell-Yan process mentioned above.  The cross section is proportional to $c_H^2$, and is thus relatively suppressed in comparison with the Drell-Yan process, especially in the large $v_\Delta$ case. 
\item[3.] {\it The weak vector boson fusion (VBF) process: }
The single production of $H_5$ occurs via the $qQ\to H_5$ process. 
The cross section is proportional to $v_\Delta^2$, so that this mode can be important in the large $v_\Delta$ case. 
\item[4.] {\it The weak vector boson associated process: }
In addition to the VBF process, $H_5$ can also be produced in association with a weak gauge boson, {\it e.g.}, $q\bar{q}'\to H_5V$.  The cross sections of such modes are proportional to $v_\Delta^2$ as for the VBF production mode.  Thus, this mode can also become important when the VBF process is important. 
\item[5.] {\it The Yukawa process: }
$H_3$ can be produced through the Yukawa interactions given in Eq.~(\ref{Yukawa1}) as the gluon fusion process for the SM Higgs boson: $gg\to H_3^0$.  
There are t-channel $H_3^\pm$ and $H_3^0$ production modes: $gb\to tH_3^-$ and $gb\to bH_3^0$. 
These production cross sections are proportional to $\tan^2\theta_H$. 
\item[6.] {\it The top quark decay: }
When $m_{H_3}$ is smaller than the top quark mass, $H_3^\pm$ can be produced from 
the top quark decay. The decay rate of the $t\to bH_3^{\pm}$ depends on $\tan^2\theta_H$. 
\end{description}

Among these production processes, channels 3 and 4 can be useful to discriminate the GM model from the others with doubly-charged Higgs bosons and to test the mass degeneracy of $H_5$.  The mDY process is also a unique feature of the GM model because the Higgs bosons $H_5$ and $H_3$ having different decay properties are produced at the same time.  In particular, when Region~II is realized, the main decay modes of these two Higgs bosons are distinctly different.  Thus, this process can be useful not only to test the mass degeneracy of $H_3$ but also to distinguish the model from the others also having doubly-charged and/or singly-charged Higgs bosons.

Let us consider the case with $m_{H_3}=150$ GeV, $\Delta m = 10$ GeV ({\it i.e.}, $m_{H_5}=140$ GeV) and $v_\Delta = 20$ GeV as an example in Region~II.  In this case, the 5-plet Higgs bosons decay into gauge boson pairs almost 100\% (the branching fractions of $H_5^0\to W^+W^-$ and $H_5^0\to ZZ$ being 67\% and 33\%, respectively).  On the other hand, $H_3^\pm$ decays to $\tau^\pm\nu$ at 66\% and $cs$ at 29\%, and $H_3^0$ decays to $b\bar{b}$ at 89\%.  We note that the branching fraction of $t\to H_3^+ b$ here is around 0.4\%.  The upper limit of the top quark decay into a charged Higgs boson and the bottom quark is 2-3\% in the case where the charged Higgs boson mass is between 80 and 160 GeV, under the assumption that the charged Higgs boson decays to $\tau\nu$ at 100\%~\cite{top_decay}.  Thus, the selected parameter set is allowed by the constraint from the top quark decays.
We consider the VBF and vector boson associated processes with the weak gauge bosons from the 5-plet decaying leptonically and the associated weak gauge bosons decaying hadronically.  Then the final states of the signal events have same-sign (SS) dileptons plus dijets and missing transverse energy ($\ell^\pm\ell^\pm jj \slashed{E}_T$) for the $H_5^{\pm\pm}$ production mode, where $\ell^\pm$ denotes collectively the light leptons $e^\pm$ and $\mu^\pm$.
The final state of the $H_5^{\pm}$ production mode includes trileptons plus dijets and missing transverse energy ($\ell^\pm\ell^\pm \ell^\mp jj \slashed{E}_T$), while that for the $H_5^0$ production mode has opposite-sign (OS) dileptons plus dijets and missing transverse energy 
($\ell^\pm\ell^\mp jj \slashed{E}_T$).  The corresponding background events for these signal events are from the $W^\pm W^\pm jj$ for the $H_5^{\pm\pm}$ production, $W^\pm Z jj$ for the $H_5^{\pm}$ production, and $t\bar{t}$, $W^\pm W^\mp jj$ and $ZZjj$ for the $H_5^{0}$ production. 

\begin{table}[h]
\begin{center}
{\renewcommand\arraystretch{1.2}
\begin{tabular}{|c||c|c|c||c|c|c||c|c|c|}\hline
&\multicolumn{3}{c||}{$\ell^\pm\ell^\pm jj \slashed{E}_T$}&\multicolumn{3}{c||}
{$\ell^\pm\ell^\pm\ell^\mp jj \slashed{E}_T$}&\multicolumn{3}{c|}
{$\ell^\pm\ell^\mp jj \slashed{E}_T$}\\\hline
Cuts&$H_5^{\pm\pm}jj$&$W^\pm W^\pm jj$&$\mathcal{S}$&
$H_5^{\pm}jj$&$W^\pm Z jj$&$\mathcal{S}$&
$H_5^{0}jj$& $t\bar{t}/VVjj$ &$\mathcal{S}$  \\\hline\hline
basic/$\Delta \eta^{jj}$/$M_T$&1.80 & 0.05 & 13.2 & 0.33 & 0.07 & 5.22 & 0.48 & 11.4 & 1.39
\\
&(5.58) & (0.12) & (23.4) & (0.98) & (0.46) & (8.17) & (1.36) & (67.4) & (1.64)
\\\hline
$b$-jet veto&-&-&-&-&-&-&0.48&1.82&3.16
\\
&-&-&-&-&-&-&(1.36)&(10.8)&(3.90)
\\\hline
\end{tabular}}
\caption{Signal and background cross sections in units of fb after each kinematic cut, along with the significance $\mathcal{S}$ based on an integrated luminosity of 100 fb$^{-1}$.  The numbers without (with) parentheses correspond to the case with a CM energy of $8$ TeV (14 TeV). 
The signal cross section includes contributions from both the VBF production and the vactor boson associated production processes.  For the $\ell^\pm\ell^\mp jj \slashed{E}_T$ events, we further impose the requirement of the $b$-jet veto for each jet to reduce the background, where the $b$-tagging efficiency is take to be 0.6~\cite{ATLAS_bjet}. }
\label{sig_and_bg}
\end{center}
\end{table}

We simulate the signal and the background event rates at the parton level for the cases where the LHC operates at the center-of-mass (CM) energy $\sqrt{s}$ of 8 TeV and 14 TeV, and impose the following basic kinematic cuts
\begin{align}
&p_T^j > 20~\text{GeV},\quad p_T^\ell > 10~\text{GeV},\quad|\eta^j| < 5,\quad |\eta^\ell| < 2.5,\quad \Delta R^{jj} > 0.4, \label{basic}
\end{align}
where $p_T^j$ and $p_T^\ell$ are the transverse momenta of the jet and the lepton, respectively, $\eta^j$ and $\eta^\ell$ are the pseudorapidities of the jet and the lepton, respectively, and $\Delta R^{jj}$ is the distance between the two jets.  To improve the significance, we further impose the cuts
\begin{align}
\Delta \eta^{jj} >3.5~~(>4.0~~\text{for}~~\ell^\pm\ell^\mp jj \slashed{E}_{T}),\quad 50<M_T <150\text{ GeV} ~,
\label{cut_add}
\end{align}
where
$M_T^2 \equiv\left[\sqrt{
M_{\text{vis}}^2+ ({\bm p}_T^{\text{vis}})^2}+| \slashed{\bm p}_T |\right]^2-
\left[{\bm p}_T^{\text{vis}}+ \slashed{\bm p}_T \right]^2 ~,~
\Delta\eta^{jj} \equiv |\eta^{j_1}-\eta^{j_2}|$. 
The results are listed in Table~\ref{sig_and_bg}.

\begin{table}[h]
\begin{center}
{\renewcommand\arraystretch{1.2}
\begin{tabular}{|c||c|c|c|c||c|c|c|c|}\hline
&\multicolumn{4}{c||}{$\ell^\pm\ell^\pm jj E_{T}\hspace{-4mm}/\hspace{4mm}$}&\multicolumn{4}{c|}
{$\ell^\pm\ell^\pm\ell^\mp jj E_{T}\hspace{-4mm}/\hspace{4mm}$} \\\hline
Cuts&$H_5^{\pm\pm}jj$&$H_5^{\pm\pm}H_3^\mp$&$W^\pm W^\pm jj$&$\mathcal{S}$&
$H_5^{\pm}jj$&$H_5^{\pm}H_3^{\mp,0}$&$W^\pm Z jj$&$\mathcal{S}$ \\\hline\hline
basic/$M_T$&3.65~(8.57) &0.71~(1.60)& 1.02~(2.20) & 18.8~(28.9) & 0.61~(1.60) &0.53~(1.21)& 1.16~(3.42) & 7.52~(11.3) \\\hline
\end{tabular}}
\caption{Signal and background cross sections in units of fb after each kinematic cut, along with the significance based on an integrated luminosity of 100 fb$^{-1}$. 
The numbers without (with) parentheses correspond to the case with a CM energy of $8$ TeV (14 TeV). }
\label{sig_and_bg2}
\end{center}
\end{table}

Next, we focus on the mDY production mode.  In order to reconstruct the masses of $H_3$ Higgs bosons, we consider hadronic decays of $H_3$, namely $H_3^\pm\to cs$ and $H_3^0\to b\bar{b}$, and 
assume leptonic decays of the weak gauge bosons from the $H_5$ decays.  Thus, the final states of the signal events from the mDY process are the same as those from the VBF and the associated processes.  Its difference from the VBF process is observed in the $\Delta\eta^{jj}$ distribution of the dijet system.  In the mDY process, the dijets in the final state come from the decay of the 3-plet Higgs boson, not the external quark jets, thus concentrating in the $\Delta\eta^{jj}\lesssim 2.5$ region.  On the other hand, the $M_T$ distributions from the mDY process and the VBF plus associated process are almost the same.  This is because the leptons plus missing transverse energy system come from the decays of $H_5$ in both processes. 
Therefore, we apply the same $M_T$ cut given in Eq.~(\ref{cut_add}) to this analysis, but not the $\Delta\eta^{jj}$ cut.  In the analysis of the mDY process, the $\ell^\pm \ell^\mp jj \slashed{E}_{T}$ signal events are overwhelmed by the huge background from the $t\bar{t}$ production.  The results are given in Table~\ref{sig_and_bg2}.

\section{Summary}

In summary, we find that $H_5^{\pm\pm}$ and $H_5^\pm$ can be detected at LHC at more than 5$\sigma$ level by using the forward jet tagging for the VBF process and the transverse mass cut on the charged leptons and missing transverse energy system if the center-of-mass energy and the luminosity are 8 TeV and 100 fb$^{-1}$, respectively.  The significance of the $H_5^0$ Higgs boson can be reached at 3$\sigma$ level by further imposing the $b$-jet veto.  We also find that the 3-plet Higgs bosons can be detected via the mDY production process.  After the $M_T$ cut, the masses of $H_3^\pm$ and $H_3^0$ can be measured from the peak in the invariant mass distribution of the dijet system.  Therefore, the respective mass degeneracies in the 5-plet Higgs bosons and the 3-plet Higgs bosons can be tested.

\begin{acknowledgments}

This research was supported in part by the National Science Council of R.O.C. under Grants Nos. NSC-100-2628-M-008-003-MY4 and NSC-101-2811-M-008-014.  The author would like to thank Shinya Kanemura for the nice organization of the workshop.
\end{acknowledgments}

\smallskip 

\end{document}